\documentstyle[12pt,epsfig]{article}
\newcommand{\lsim}{\mathrel{\lower4pt\hbox{$\sim$}}
\hskip-12.5pt\raise1.6pt\hbox{$<$}\;}

\newcommand{\gsim}{\mathrel{\lower4pt\hbox{$\sim$}}
\hskip-12.5pt\raise1.6pt\hbox{$>$}\;}

\def\g2{{$(g-2)$}}

\def\bc{\begin{center} }
\def\ec{\end{center} }

\def\beq{\begin{equation}}
\def\eeq#1{\label{#1}\end{equation}}
\def\eeqn{\end{equation}}

\def\beqa{\begin{eqnarray}}
\def\eeqa#1{\label{#1}\end{eqnarray}}
\def\eeqan{\end{eqnarray}}

\let\bar=\overbar

\def\Dslash{\not{\hbox{\kern-4pt $D$}}}
\def\dslash{\not{\hbox{\kern-2pt $\del$}}}

\def\msb{{\bar{\ssstyle M \kern -1pt S}}}

\def\g2{{$(g-2)$}}

\def\bea{\begin{eqnarray}}
\def\eea{\end{eqnarray}}
\def\bc{\begin{center} }
\def\ec{\end{center} }
\def\beq{\begin{equation}}
\def\eeq#1{\label{#1}\end{equation}}
\def\eeqn{\end{equation}}
\begin{document}
\title{ Recent Progress on the BNL Muon $(g-2)$ Experiment\footnote{Invited
Talk at KAON2001 - Pisa, 12-17 June 2001.  (corresponding author: 
roberts@bu.edu)} }
\maketitle
\noindent{ {B. Lee Roberts$^1$, 
H.N. Brown$^2$, G. Bunce$^2$, R.M. Carey$^1$,
P. Cushman$^{9}$, G.T. Danby$^2$,
P.T. Debevec$^7$, M. Deile$^{11}$, H. Deng$^{11}$, W. Deninger$^7$,
S.K. Dhawan$^{11}$, V.P. Druzhinin$^3$,
L. Duong$^{9}$, E. Efstathiadis$^1$, F.J.M. Farley$^{11}$,
G.V. Fedotovich$^3$, S. Giron$^{9}$,
F. Gray$^7$, D. Grigoriev$^3$,
 M. Grosse-Perdekamp$^{11}$,
A. Grossmann$^6$, M.F. Hare$^1$,
D.W. Hertzog$^7$, V.W. Hughes$^{11}$, M. Iwasaki$^{10}$,
K. Jungmann$^6$, D. Kawall$^{11}$,
M. Kawamura$^{10}$, B.I. Khazin$^3$, J. Kindem$^{9}$, F. Krienen$^1$,
I. Kronkvist$^{9}$, R. Larsen$^2$,
Y.Y. Lee$^2$, I. Logashenko$^{1}$, R. McNabb$^{9}$, W. Meng$^2$,
J. Mi$^2$, J.P. Miller$^1$, W.M. Morse$^2$,
D. Nikas$^2$, C.J.G. Onderwater$^7$, Y. Orlov$^4$,
C.S. \"{O}zben$^2$,
J.M. Paley$^1$, C. Polly$^7$,
J. Pretz$^{11}$, R. Prigl$^2$, G. zu Putlitz$^6$, 
S.I. Redin$^{11}$,  O. Rind$^1$, N. Ryskulov$^3$,
S. Sedykh$^7$, Y.K. Semertzidis$^2$, Yu.M. Shatunov$^3$,
E.P. Sichtermann$^{11}$, E. Solodov$^3$,
M. Sossong$^7$, A. Steinmetz$^{11}$, L.R. Sulak$^1$,
C. Timmermans$^{9}$, A. Trofimov$^1$,
D. Urner$^7$, P. von Walter$^6$, D. Warburton$^2$, D. Winn$^5$,
A. Yamamoto$^8$, D. Zimmerman$^{9}$,}\\
{\em
$^1$Department of Physics, Boston University, Boston, MA 02215, USA,
$~^2$Brookhaven National Laboratory, Upton, NY 11973, USA,
$~^3$Budker Institute of Nuclear Physics, Novosibirsk, Russia,
$~^4$Newman Laboratory, Cornell University, Ithaca, NY 14853, USA,
$~^5$Fairfield University, Fairfield, CT 06430, USA,
$~^6$Physikalisches Institut der Universit\"{a}t Heidelberg, 69120
Heidelberg, Germany,
$~^7$University of Illinois at Urbana-Champaign, IL
61801, USA,
$~^8$KEK, Tsukuba, Ibaraki
305-0801, Japan,
$~^{9}$University of Minnesota, Minneapolis, MN
55455, USA,
$~^{10}$Tokyo Institute of Technology, Tokyo, Japan,
$~^{11}$Yale University, New Haven, CT 06520, USA
}
}
%\maketitle
%\baselineskip=11.6pt
\begin{abstract}
The status of the muon \g2 experiment at the Brookhaven AGS is reviewed.
An accuracy of 1.3 ppm on the $\mu^+$ anomalous magnetic moment
has been achieved and published. This result differs with the 
standard model prediction by about 
2.5 standard deviations.
A data sample with approximately
seven times as much data is being analyzed, with a result
expected in early 2001. 
\end{abstract}
\baselineskip=14pt
\section{Introduction}

Measurements of the $g$-factors of elementary particles 
have been
intimately tied to the development of our
understanding of subatomic physics.
The proportionality constant, $g$, relates the spin and magnetic
moment of an elementary particle through the relationship
\beq
\vec \mu_s = g( {e \over 2 m }) \vec S, \qquad {\rm where} \qquad
a = {(g-2) \over 2}
\eeqn
is the anomalous part of the magnetic moment.
In the Dirac theory, 
$g = 2.$ 
The muon anomalous moment is
dominated by the lowest order radiative correction $\alpha \over 2\pi$,
which was first calculated in 1948 by Schwinger.\cite{schwinger}
QED calculations of the electron and muon anomalies
have been carried out to eight-order (with an estimate of 
tenth-order) by
Kinoshita and others.\cite{kinhughes}

\begin{figure}[ht]
%\begin{center}
 \vspace{1.6cm}
\includegraphics{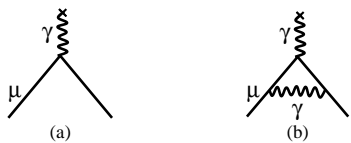}
\caption{\it
The Feynman graphs for (a) $g=2$; and (b) the lowest
order radiative correction (Schwinger term).
\label{fig:geq2sch}}
%\end{center}
\end{figure}

The electron anomalous moment is now measured to an experimental accuracy
of a few parts per billion,
and is well described by QED
calculations.\cite{kinhughes}
To the level of measurement, only photons and electrons
contribute, and $a \simeq 1 \times 10^{-3}$.
There
is no evidence to date, either from $g$-factor measurements or $e^+e^-$
scattering, to indicate that the electron has any internal structure.

While the $g$-factor of the electron has provided a testing ground for QED,
the anomalous magnetic moment of the muon has provided an even richer source of
information,  since the contribution
of heavy virtual particles to the 
anomaly scales as the mass of the lepton squared.
In a series of three elegant experiments at 
CERN,\cite{cern3}
virtual
muons and hadrons have been shown to
contribute at measurable levels.  

The CERN experiments measured $\mu_{\mu} = [0.001$ $165$ $9230(85)]$
$(e\hbar/2
m_{\mu})$, a precision of $\pm 7.3$ parts per million
(ppm) for $a_{\mu}$.  This result
tested QED to a high level, and showed for the first
time the contribution of virtual hadrons to the magnetic moment of a lepton,
the sensitivity was not sufficient to observe the predicted
electroweak contribution of 1.3 ppm from virtual $W^{\pm}$ and
$Z^0$ gauge bosons.
The goal of our experiment is an overall accuracy
of $\pm 0.35$
ppm which allows
sufficient sensitivity to measure the electroweak contribution,
as well as
to search for physics beyond the
standard model.

\begin{figure}[htb]
 \vspace{2.1cm}
\includegraphics{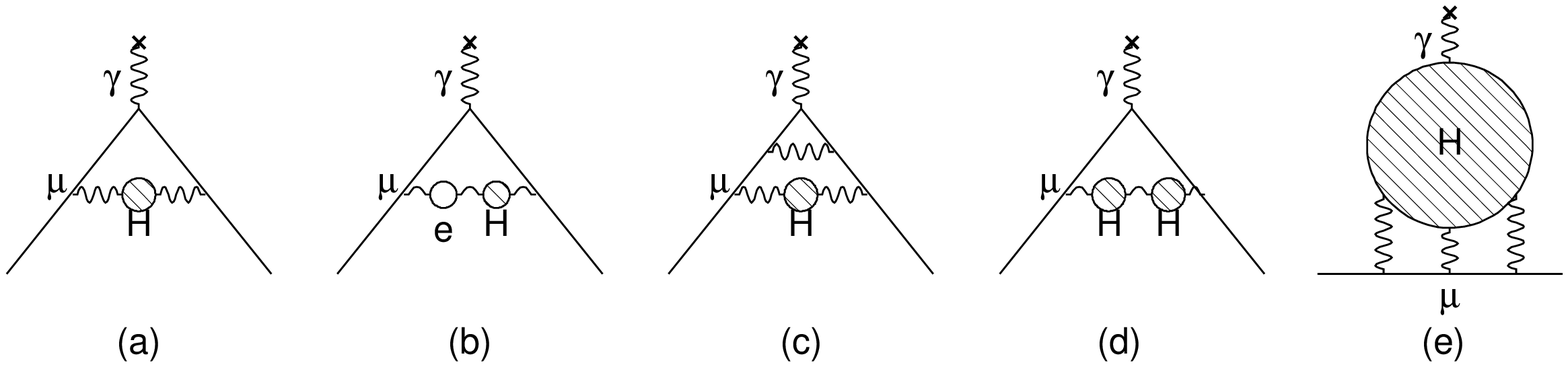}
\caption{\it Hadronic contributions to the muon anomalous moment. 
In these diagrams, H refers to a loop with hadrons (quarks).
(e) shows the hadronic light-by-light contribution discussed by 
Prades at this meeting.\cite{prades}
\label{fig:had}}
\end{figure}

The standard model 
theory has been reviewed at this 
conference by Prades.\cite{prades}  
The theoretical value of $a_{\mu}$ consists of contributions from
QED, virtual
hadrons,
and virtual electroweak gauge 
bosons.\cite{newphys2}
The theoretical uncertainty is dominated by the hadronic contribution, which
must be calculated using data from 
$e^+e^- \rightarrow {\rm hadrons}$ along
with a dispersion relation, or in the case of
the hadronic light-by-light term, from a model calculation.
The most important diagrams are shown in Fig. \ref{fig:had}.
The largest contribution (and the largest uncertainty) comes from 
the hadronic vacuum polarization diagram (Fig. \ref{fig:had}(a)).

The one-loop electroweak contributions to $a_{\mu}$ 
have been available for some
time, and now  higher order calculations
which include both fermionic
and bosonic
two-loop terms are available, with the next order leading logs also
evaluated.

One of the main motivations for our measurement was to confront the 
standard model, and to search for possible contributions from non-standard
model physics such as supersymmetry, muon or $W$-gauge-boson substructure.
Theoretical interest in possible non-standard model contributions to
the muon \g2 value has risen substantially in the past five years, and
a great deal has been written about possible contributions to 
the muon \g2 value from non-standard model 
physics.\cite{newphys2} 

Just as proton
substructure produces a $g$-value which is not equal to two, muon
(or $W$)
substructure would also contribute to the anomalous moment, the critical
issue being the scale of the substructure.
A standard model
value for \g2 at the 0.35 ppm level would restrict the substructure scale
to around 5 TeV.  If 
leptoquarks exist, they too could contribute to the non-standard model
value of \g2.
The muon \g2 obtains its sensitivity to
$W$ substructure and anomalous gauge couplings through
 the  $WW\gamma$  triple gauge vertex in the single loop $W$ contribution.

\begin{figure}[htb]
\vspace{2.2cm}
\includegraphics{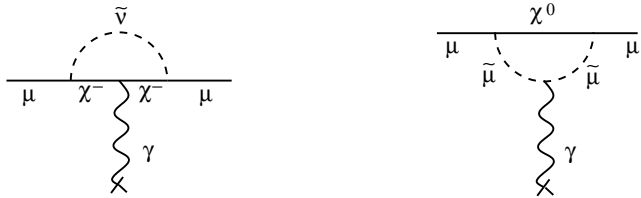}
\caption{\it The lowest order supersymmetric contributions to \g2.
\label{fig:susy}}
\end{figure}

Supersymmetry has become a serious candidate for physics beyond the standard
model.  There is a large sensitivity to almost any supersymmetric model
with large $\tan \beta$.\cite{newphys2}q
The SUSY contribution is shown in 
Fig. \ref{fig:susy}.
In the case of large $\tan \beta$, the chargino ($\chi^-$) diagram dominates
the contribution to \g2 from SUSY, and is given by\cite{newphys2}
\beq
 a_{\mu}({\rm SUSY}) 
\simeq 
{\alpha \over 8 \pi \sin^2 \theta_W} {m_{\mu}^2 \over \tilde m^2} \tan \beta 
\simeq { 140 \times 10^{-11} } \  ({100\ {\rm GeV} \over \tilde m})^2
\tan \beta ,
\eeq{eq:susy}
where $\tilde m$ is the largest mass in the loop.
The goal of E821 is to reach a precision of
$\pm 40\times 10^{-11}$ ($\pm 0.35$ ppm), so 
the factor of
140 in Eq. \ref{eq:susy} corresponds to 1.2  ppm in $a_{\mu}$.  
For $\tilde m= 750$ GeV 
and $\tan \beta = 40$,
$a_{\mu}({\rm SUSY}) = 100\times 10^{-11} $, a contribution which is
2.5 times larger than 
the sensitivity we hope to achieve.  For $\tilde m = 500$ GeV, the effect
is $224\times 10^{-11} $ or 5.6 ppm.

\section{The Experimental Technique}

The experimental technique is a refinement of the technique used in the 
third CERN experiment,\cite{cern3} 
with the addition of direct injection
of a muon beam into the storage ring.
A superconducting magnetic storage ring 
with a vertical magnetic field of 1.45 T, central orbit radius of 711.2 cm,
and central momentum of 3.1 GeV stores a bunch of muons provided by the 
AGS. Vertical focusing is provided by electrostatic quadrupoles 
which are placed in the ring with four-fold symmetry.  The injected beam
is kicked onto a stable orbit by a fast kicker system which uses
a current distribution to provide the $\sim0.1$ Tm kick needed to store
the beam. The residual magnetic field from the fast kicker has been measured
using the Faraday effect, and it was found to contribute less that 0.1 ppm 
to the integrated $B dl$ seen by the muons for times greater than 
20 $\mu$s after injection.

A charged particle moving transverse to 
a uniform magnetic field will go in a circle
with the orbital cyclotron frequency
\beq
\omega_c = {(eB) \over  (m \gamma)}.
\eeqn
The spin precession frequency in this same magnetic field is given by
\beq
\omega_s = {geB\over 2m} + (1-\gamma){eB\over m \gamma}. 
\eeqn
Thus the spin vector of
a charged particle moving transverse to
a uniform magnetic field will precess
relative to the momentum vector
with a frequency $\omega_a$, which is given by the difference between the
orbital cyclotron frequency $\omega_c$ and the spin precession frequency
$\omega_s$.  This frequency is
\beq
\omega_a = \omega_s - \omega_c = { e \over m} a_{\mu} B,
\eeq{eq:omegaa}
 where
$\omega_a$ is directly proportional to the anomalous
moment and is
independent of the particle's momentum.

  Vertical focusing
must be provided to keep the muon beam stored, which can be accomplished
with magnetic
multipoles, or with an electrostatic quadrupole field.  However,
if magnetic multipoles
are used, it is difficult to determine
 the average $B$ field to the accuracy needed
for a precision measurement of $a_{\mu}$.
In a region in which both
magnetic and electric fields are present, the relativistic formula for the
precession is given by
\beq
 \vec \omega_a = {d \Theta_R \over dt} = -{e \over m }\left[ a_{\mu} \vec B -
\left( a_{\mu}- {1 \over \gamma^2 - 1}\right)
\vec \beta \times \vec E \right],
\eeq{eq:bmt}
where $\Theta_R$
is the angle between the muon spin
direction in its rest frame and the muon velocity direction in the laboratory
frame.  The other quantities refer to the laboratory frame.
If the muon beam has the "magic" value of $\gamma =29.3$, then the coefficient
of the $\vec \beta \times \vec E$ term is zero, and the electric field
does not cause spin precession.
Thus the precession of the spin relative to the momentum is
determined entirely by the magnetic field, and one can use electrostatic
quadrupoles for vertical focusing.
Because the muon's lifetime is relatively long, and because muons are
produced fully-polarized along their direction of motion in pion decay at
rest, it is possible to produce a beam of  polarized
muons.  With our kicker we store $\sim 10^4$ per fill of the storage ring.

 In the three-body decay
$\mu^+ \rightarrow e^+ \bar \nu_{\mu} \nu_e $,
the highest energy positrons are preferentially emitted
parallel to the muon spin direction in the muon rest frame.
When emitted parallel to the muon momentum,
the highest energy electrons in the muon rest frame are Lorentz boosted to
become the highest energy
electrons in the lab frame. Therefore, the number of high energy electrons
is a maximum when the muon spin is parallel to the momentum, and
a minimum when it is anti-parallel, thus making
it possible to measure the spin (or anomalous) precession frequency
by counting high energy electrons as a function of time.  This time
spectrum will show the muon lifetime modulated by the spin precession
frequency.  In a perfect experiment, the positron time spectrum would
be given by
\beq
N(t)=N_0e^{-t / \gamma \tau}(1+A\cos{(\omega_a t+\phi)}).
\eeq{eq:fiveparam}
The real experiment has pulse pileup, muons lost from the
storage ring other than by decay, and because the detector acceptance
depends on the
radial position of the muon decay, the coherent motion of the beam
in the storage ring modulates the time spectrum with the coherent
betatron frequency.  In the four independent analyses of the 1999 data
set, pileup was either fit directly, or subtracted from the data set.
A picture of the pileup subtracted data and a ten parameter fit to it  
is given
in Fig. \ref{fig:wiggle} below.

\begin{figure}[htb]
\vspace{6.5cm}
\includegraphics{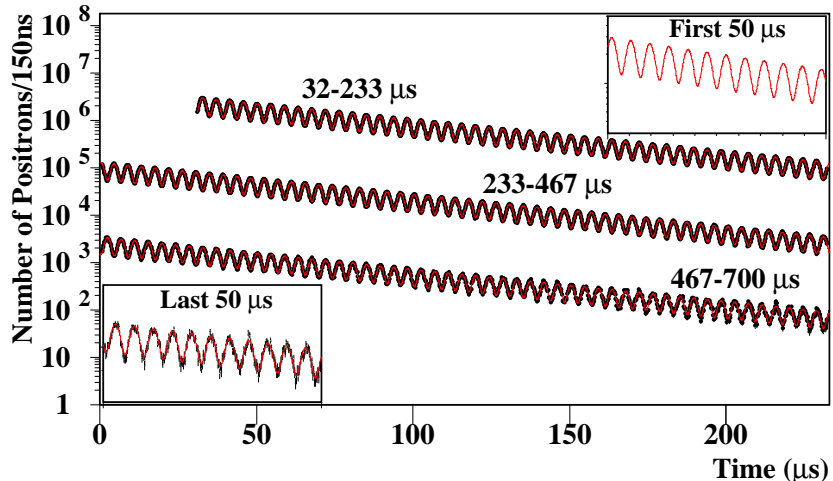}
\caption{\it The positron time spectrum from the 1999 data set
with an energy threshold of 2.0 GeV.
The data and fit to a 10 parameter function are shown, where
$\chi^2/\nu$ for the fit is 3818/3799.  There are $0.95\times 10^9$
events in the histogram.
\label{fig:wiggle}}
\end{figure}

Equation \ref{eq:bmt} gives the principal elements needed to obtain
a value for the muon anomaly.  One needs to measure the muon 
frequency $\omega_a$ as well as the magnetic field weighted over the muon
distribution.  The field is measured with NMR techniques, in a 
four-step 
procedure.  During data collection, the field is monitored with
377 fixed NMR probes which are outside of the beam vacuum chamber.  
Every few days an NMR mapping trolley is used to
map the field inside of the storage region, which calibrates the
fixed probes.  Before and after the running period, the NMR probes in
the trolley are cross calibrated with a calibration probe located
at one point in azimuth, which 
plunges into the storage region to
measure the field at the position of
each of the 17 trolley probes.  This plunging probe, is then 
calibrated with a special probe which has a spherical water sample,
thus giving us the spin rotation frequency of a free proton, $\omega_p$, in
our magnetic field.  We compute the ratio $R=\omega_a/\omega_p$,
and the anomaly is given by
$ a_{\mu} = R/(\lambda -R)$. 
The constant
$\lambda=\mu_{\mu}/\mu_p$, the ratio of the magnetic 
moments of the muon and proton, is known independently from
other experiments.

The analysis was performed ``blind'' meaning that arbitrary offsets were
put on the two frequencies $\omega_a$ and $\omega_p$ during the analysis,
so that it was impossible to determine the value of without knowing these
two offsets. Two independent analyses of $\omega_p$ 
and four of $\omega_a$  were performed. Only after the separate 
analyses of these frequencies were consistent and well studied,
were the offsets revealed, and the value of $a_{\mu}$ computed.

\begin{figure}[htb]
\vspace{5.5cm}
\includegraphics{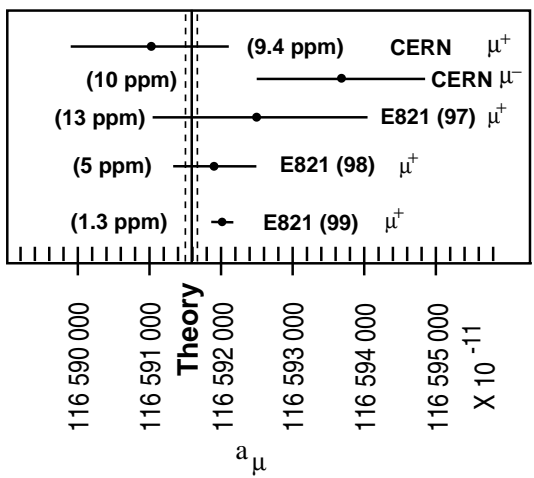}
\caption{\it Experimental and theoretical values for $a_{\mu}$.
\label{fig:value}}
\end{figure}

The value obtained\cite{brown2} was
$a_{\mu^+}=11\ 659\ 202(14)(6) \times 10^{-10}~(1.3~\mathrm{ppm})$.
The standard model value used for comparison 
was\cite{newphys2}
$a_{\mu}=11\ 659\ 159.6(6.7)\times 10^{-10}~(0.66~\mathrm{ppm})$.
The weighted averaged of the experimental values gives
a difference from the standard model of
$a_{\mu}(exp)-a_{\mu}(th)=+43(16)\times 10^{-10}$.  The individual
measurements are shown graphically in Fig. \ref{fig:value}.

While this difference with the standard model is quite interesting, it
is far from definitive.  As discussed at this conference by Prades,
the theoretical value is undergoing detailed study by a number of
people.  As additional data from Novosibirsk, $Da\phi ne$ and BES
become available, along with the full $\tau$-decay data set from
LEP and CLEO, our knowledge of the theoretical hadronic contribution
will improve further.

The experimental value is also a work in progress.
We collected about 4 billion positrons in our 2000 run,
and about 3 billion electrons in our 2001 run.  Thus we expect our
statistical error to improve by about $\sqrt 7$.  We are constantly 
improving our understanding of the systematic errors, and believe that
the final total systematic error should be about 0.3 ppm.  Recently,
scientific approval for an additional data collection period was 
given by the Laboratory, but funding will have to be found if we
are to collect these additional data.  If we obtain the additional
data, we expect to reach a statistical error of about 0.33 ppm.

Much progress has been made since our collaboration began almost two
decades ago.  We have solved many interesting technical issues, 
and have obtained an
answer at the part per million level.  This new result presented us
with a surprise which has been received with widespread interest.
Our collaboration is working hard to finish the analysis of the 
additional data sets in order to clarify whether this potential signal
for new physics will remain at the end of the day.  Stay tuned.

%

%\begin{table}[t]
%  \centering
%  \caption{ \it Example of a table.
%    }
%  \vskip 0.1 in
%  \begin{tabular}{|l|c|c|} \hline
%    &  experiment & simulation \\
%    \hline
%    \hline
%    side-on   & $ (4.81 \pm 0.06)\%$ $E^{-\frac{1}{2}}$   & 
%    $ (4.70 \pm 0.05)\%$ $E^{-\frac{1}{2}}$                    \\
%   head-on   & $ (4.7 \pm 0.1)\%$ $E^{-\frac{1}{2}}$   $ + (3.4 \pm 0.6)\%$  &
%    $ (4.6 \pm 0.3)\%$ $E^{-\frac{1}{2}}$   $ + (3.8 \pm 1.3)\%$ \\
%    \hline
%  \end{tabular}
%  \label{extab}
%\end{table}
%

%
\end{document}